# Three classical Cepheid variable stars in the nuclear bulge of the Milky Way

Noriyuki Matsunaga[1*], Takahiro Kawadu[2], Shogo Nishiyama[3],

Takahiro Nagayama[4], Naoto Kobayashi[1,5], Motohide Tamura[3],

Giuseppe Bono[6,7], Michael W. Feast[8,9] & Tetsuya Nagata[2]

[1] Kiso Observatory, Institute of Astronomy, University of Tokyo, 10762-30 Mitake, Kiso-machi, Kiso-gun, Nagano 397-0101, Japan.

[2] Department of Astronomy, Kyoto University, Kitashirakawa-Oiwake-cho, Sakyo-ku, Kyoto 606-8502, Japan.

[3] National Astronomical Observatory of Japan, 2-21-1 Osawa, Mitaka, Tokyo 181-8588, Japan.

[4] Department of Astrophysics, Nagoya University, Furo-cho, Chikusa-ku, Aichi 464-8602, Japan.

[5] Institute of Astronomy, University of Tokyo, 2-21-1 Osawa, Mitaka, Tokyo 181-0015, Japan.

[6] Departmento di Fisica, Universitá di Rome Tor Vergata, Via della Ricerca Scientifica 1, 00133 Rome, Italy.

[7] INAF—Osservatorio Astronomico di Roma, Via Frascati 33, 00040 Monte Porzio Catone, Italy.

[8] Astrophysics, Cosmology and Gravity Centre, Astronomy Department, University of Cape Town, 7701 Rondebosch, South Africa.

[9] South African Astronomical Observatory, PO Box 9, 7935 Observatory, South Africa.

---

* matsunaga@ioa.s.u-tokyo.ac.jp

**The nuclear bulge is a region with a radius of about 200 parsecs around the centre of the Milky Way[1]. It contains stars with ages[2-4] ranging from a few million years to over a billion years, yet its star-formation history and the triggering process for star formation remain to be resolved. Recently, episodic star formation, powered by changes in the gas content, has been suggested[5]. Classical Cepheid variable stars have pulsation periods that decrease with increasing age[6], so it is possible to probe the star-formation history on the basis of the distribution of their periods[7,8]. Here we report the presence of three classical Cepheids in the nuclear bulge with pulsation periods of approximately 20 days, within 40 parsecs (projected distance) of the central black hole. No Cepheids with longer or shorter periods were found. We infer that there was a period about 25 million years ago, and possibly lasting until recently, in which star formation increased relative to the period of 30—70 million years ago.**

We conducted a near-infrared survey for the 0.33° by 0.5° area around the Galactic Centre[9], where no classical Cepheids were known before[10,11]. Approximately 90 time-series images were collected in each of the $J$ (1.25 μm), $H$ (1.63 μm) and $K_S$ (2.14 μm) wave-passbands during eight years between 2001 and 2008. We discovered 45 variable stars with pulsation periods shorter than 60 days, among which we determined three to be classical Cepheids (Table 1, Fig. 1 and Supplementary Fig. 1) on the basis of their light curves and other observed properties (Supplementary Information). Both their extinctions and distances can be determined[9] using the period-luminosity relations in the three passbands[12]. Their derived distances (Table 1) agree closely with each other. The mean distance is 7.9 kiloparsecs (kpc), with standard error of the mean ±0.2 kpc, although there remains a systematic error of ±0.3 kpc, of which the dominant source is an error in the total-to-selective extinction ratio (that is, the coefficient used to convert a colour excess into an extinction). This result from a fundamental distance indicator agrees with the distance to the Galactic Centre derived from orbits round the central black hole[13] (8.28±0.33 kpc) amongst other estimates[9,14-16]. We also note that the Cepheids are located close to the plane containing the central black hole ($b$=−0.05°), well within 10 parsec (pc) in projected distance, which is consistent with their being within the thin disk-like structure of the nuclear bulge[1,5]. These results clearly locate the three objects inside the nuclear bulge.

All of our classical Cepheids have periods close to 20 days. Figure 2 compares their period distribution with that of known Cepheids in a wide region around the Sun[17]. Despite the small number of the objects found, the histogram contrasts our targets with the absence of shorter-period Cepheids in the nuclear bulge. Although the older, shorter-period Cepheids are expected to be the fainter ones[6], our survey would have detected those with periods longer than 5 days if they existed (Supplementary Fig. 2). Our sample shows a concentration in the period range where the proportion of Cepheids is generally small; such a period distribution has not been seen in nearby galaxies[18]. The period distribution can be affected by metallicity[18], but the metallicities of young objects in the nuclear bulge have been reported to be approximately solar[19]. The 20-day Cepheids have an age of 25±5 million years (Myr), where the error is derived from the standard deviation of the period-age relation[6]. The uncertainty in the metallicity has a smaller effect on the age estimate than the above error[6]. All three Cepheids found in the nuclear bulge are approximately the same age. We discuss the possibility that the two stars projected close to each other might have formed within a cluster, although this is not probable, in the Supplementary Information.

We can estimate the star-formation rate at about 25 Myr ago by assuming an initial mass function[20] and the lifetime spent by the Cepheid inside the instability strip[21]. To allow comparison with previous work, we take into account the survey area and the region hidden by extreme interstellar extinction. We find that the star-formation rate was $0.075^{+0.15}_{-0.05}$ solar masses per year in the entire nuclear bulge 20—30 Myr ago. The uncertainty comes from Poisson noise, the uncertainty in the Cepheid lifetime, and the uncertainty in the ratio of the effective survey area to the entire nuclear bulge (Supplementary Information). On the other hand, the absence of shorter-period Cepheids leads to 0.02 solar masses per year as a 1σ upper limit on the star-formation rate for 30—70 Myr ago. The Poisson noise remains the dominant uncertainty in our discussions. We can ignore the uncertainty in the effective area of our survey, a factor of two, when we compare the two star-formation rates. If the star-formation rate remained as high as the value obtained with the 20—30-Myr-old Cepheids, the probability of finding no 30—70-Myr-old Cepheid is low. Thus we conclude that the change in the star-formation rate between 20 and 70 Myr ago is significant at the level of 2σ, assuming the Poisson statistics.

A recent investigation suggested that the star-formation rate was low a few tens of millions of years ago and then increased to a peak at about 0.1 Myr ago, followed by a decline in very recent times[5]. However, the tracers used give only a rough time scale for the range 1—100 Myr ago. Our estimates have much higher time resolution for the 20—70-Myr range and indicate an increase in star-formation rate within this period. These results are illustrated in Fig. 3, which also shows a scenario of continuous star formation[4] (the value of ref. 4 has also been corrected to our scale by taking into account the mass of the entire nuclear bulge, which is $(1.4\pm0.6)\times10^9$ solar masses[1]).

It is of interest to consider, in the context of galactic structure and evolution, how and why such time variations in star formation occurred. Episodic star formation has been suggested in some of the so-called pseudobulges[22,23], the central regions of a few barred spiral galaxies, possibly growing with bar-driven gas inflow[24]. Likewise, our result suggests that episodic star formation on a short time scale of about 25 Myr occurred in the nuclear bulge, which some authors claim to be a pseudobulge[25]. The time scale is comparable with that of the cyclic gas accumulation predicted for the central part of the Milky Way[26].

**Acknowledgements** We thank the IRSF/SIRIUS team in Nagoya University, the National Astronomical Observatory of Japan, Kyoto University and the University of Tokyo, and the staff of the South African Astronomical Observatory (SAAO) for their support during our near-IR observations. This work was supported by Grants-in-Aid for Scientific Research from the Japan Society for the Promotion of Science (JSPS). M.W.F. acknowledges the support from National Research Foundation (NRF) of South Africa.

**Author Contributions** N.M. led the programme, carried out the analysis and wrote most of the text. T.K., S.N. and Ta.N. as well as N.M. carried out the monitoring observations at the IRSF. S.N., N.K., M.T., G.B., M.W.F. and Te.N. took part in the discussions especially on the contexts which are close to their own research fields such as evolution of galaxies, star formation and stellar pulsation. All authors contributed to the writing of the paper.

**Table 1 | Catalogue of the classical Cepheids in the nuclear bulge**

RA, right ascension; Dec. declination (J2000.0); $l$ and $b$ are the galactic coordinates; $\Delta_l$ and $\Delta_b$ are projected distances of the Cepheids from the central black hole at 8 kpc; $J$, $H$, and $K_s$ are the mean intensity magnitudes in those wave-passbands; $P$ are the periods. The standard errors for the estimates of the true distance modulus $\mu_0$ and the $K_s$-band foreground extinction $A_{K_s}$ are ±0.13 and ±0.08 mag, respectively (see Supplementary Information).

| Star | RA | Dec. | $l$ (deg) | $b$ (deg) | $\Delta_l$ (pc) | $\Delta_b$ (pc) | $J$ (mag) | $H$ (mag) | $K_s$ (mag) | $P$ (days) | $\mu_0$ (mag) | $A_{K_s}$ (mag) |
|---|---|---|---|---|---|---|---|---|---|---|---|---|
| a | 17:46:06.01 | −28:46:55.1 | +0.1864 | −0.0095 | +33.9 | +6.5 | 15.63 | 12.04 | 10.18 | 23.54 | 14.55 | 2.46 |
| b | 17:45:32.27 | −29:02:55.2 | −0.1054 | −0.0433 | −6.9 | +0.4 | 15.42 | 12.00 | 10.17 | 19.96 | 14.49 | 2.35 |
| c | 17:45:30.89 | −29:03:10.5 | −0.1116 | −0.0412 | −7.8 | +0.7 | 16.36 | 12.44 | 10.35 | 22.76 | 14.42 | 2.74 |

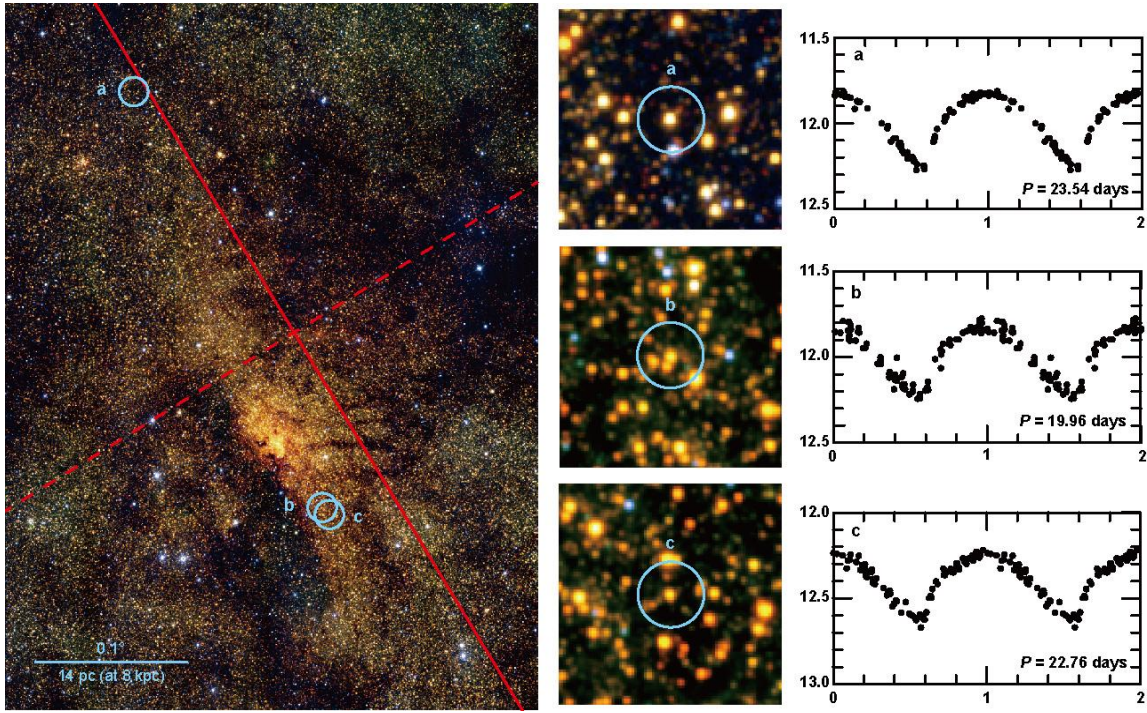

**Figure1 | The classical Cepheids discovered in the nuclear bulge.**

The left panel shows the observed field of 0.33° by 0.5°. This false-colour image is composed of images, a subset of the data we used in analysis, in three passbands: $J$ (1.25 μm), $H$ (1.63 μm) and $K_S$ (2.14 μm). The data were taken with the Infrared Survey Facility (IRSF) 1.4-m telescope and the SIRIUS near-infrared camera located at the South African Astronomical Observatory. North is up and east is left. The Galactic plane (that is, $b=0°$) is indicated by the solid red line running from the top-left to the bottom-right, while the dashed red line shows the axis of $l=0°$. An angle of 0.1° corresponds to a scale of 14 pc at the distance of the Galactic nuclear bulge (about 8 kpc), as indicated in the image.

The central black hole is located at the heart of the most crowded region in the chart ($l=-0.056°$, $b=-0.046°$). Our survey field includes two famous clusters: Quintuplet ($l=-0.16°$, $b=+0.06°$) and Arches ($l=-0.12°$, $b=-0.02°$). The three panels in the middle column are close-ups, around the circles labelled a, b and c in the left panel, covering a 40-arcsecond square around each object in Table 1. Again, north is up and east is left. The right panels show the light curves in the $H$ band for the stars labelled a, b and c. The variations of the $H$-band magnitude are folded according to the periods $P$ indicated, and plotted against the phase. The three light curves, showing a clear resemblance to each other, have the typical variations of classical Cepheids.

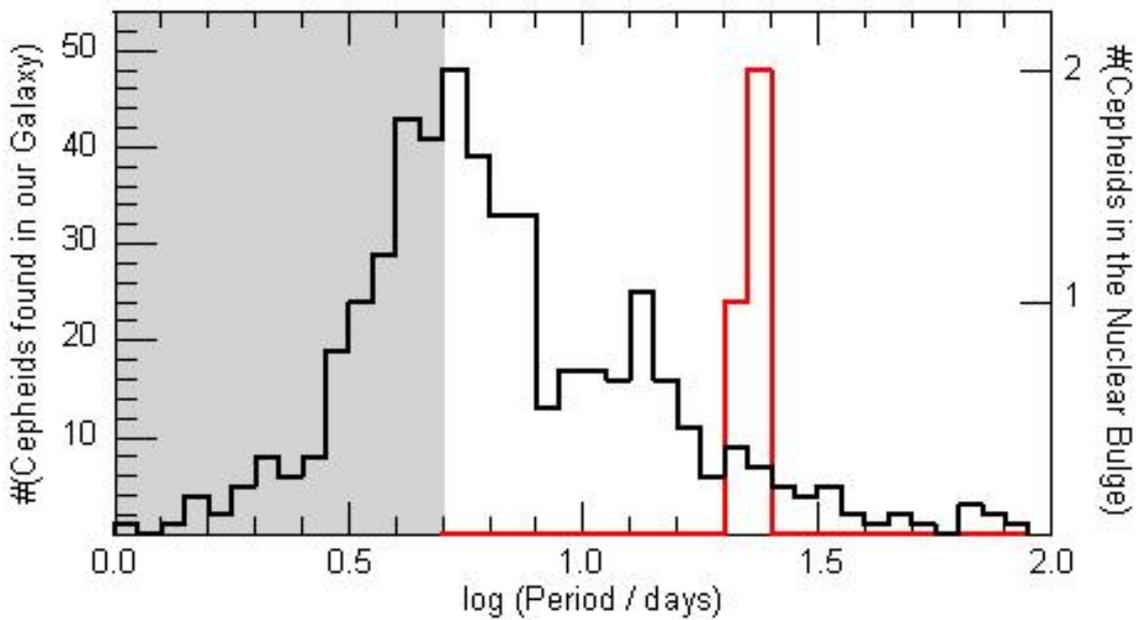

**Figure 2 | Period distribution of the nuclear bulge Cepheids and those previously found in the Milky Way.**

The period distribution of the three Cepheids in the nuclear bulge (red) is compared with that of the previously known Cepheids in the Milky Way[17] (black). The grey region indicates the period range for which our survey could not reach classical Cepheids in the nuclear bulge with typical extinction of $A_{Ks}<3$ mag, even if they were there (see Supplementary Information).

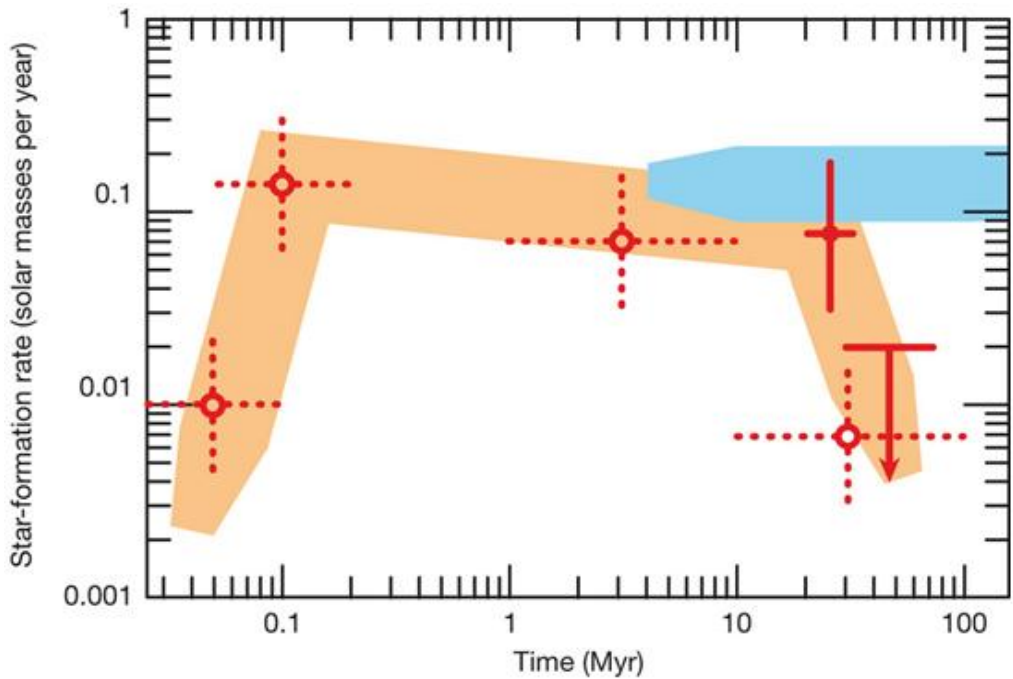

**Figure3 | Star-formation history in the nuclear bulge during the last hundred million years.**

The star-formation rates we derived are indicated by the point at 25 Myr and the arrow at about 50 Myr. The vertical error bar for the point at 25 Myr corresponds to a factor of 2.5, taking errors like the Poisson noise of the Cepheid number into account. The vertical error bar for the arrow at about 50 Myr indicates the 1σ upper limit. The uncertainty in the correction for our effective survey coverage (a factor of 2) is not included here, in order to illustrate the difference between these estimates, but it should be considered when the star-formation rates are compared with other data below. The horizontal error bars indicate the durations for the corresponding estimates. More details on these error bars are given in the Supplementary Information. The star-formation rates given in ref. 5 are indicated by open circles, and the dashed horizontal bars show the approximate uncertainty in the age for each tracer. The errors of the star-formation rates in ref. 5 were not given, but they cannot be smaller than a factor of two, considering the uncertainty in the age estimate of young stellar objects. The orange stripe indicates our scenario of the episodic star formation, combining our results and those of ref. 5, while the blue stripe indicates the continuous star-formation scenario suggested in ref. 4.

# Supplementary Information

**1. Identification of classical Cepheids:**

We identified three classical Cepheids amongst 45 discovered short-period variables as follows. Our sample includes classical Cepheids, type II Cepheids, eclipsing binaries and other types of pulsating stars. In particular, type II Cepheids have similar periods to classical Cepheids[27]. They are old and evolved from low-mass stars, ~1 solar mass. They also have a period-luminosity relation, which is fainter than that of classical Cepheids by 1.5—2 mag[27,28]. In order to select classical Cepheids, the shapes of the light curves, the distances and extinctions derived with the period-luminosity relations were considered. Because of the difference between the period-luminosity relations, the estimated distances are substantially different depending on whether the variables are classical or type II Cepheids. We make use of the period-luminosity relation of the classical Cepheids with the parallaxes measured by the *Hubble Space Telescope* in the solar neighbourhood[12] and the relation of type II Cepheids in the Large Magellanic Cloud[28]. The reddening law, i.e. the wavelength dependence of extinction, is adopted from the result obtained with the same instrument[29]. These, together with the apparent distance moduli derived in the different wavebands allow us to estimate both the extinction and the true distance modulus.

We conclude that three objects are classical Cepheids (Table 1). Fig. 1 shows the folded light curves of the three, which follow the typical variations expected for classical Cepheids with the corresponding periods. Their mean distance, 7.9 ± 0.2 (s.e.m.) ± 0.3 (systematic) kpc, and their estimated extinctions, $A_{K_s}$ ~2.5 mag, are consistent with other objects in the nuclear bulge[9]. The estimated standard errors of individual distance moduli ($\mu_0$) and extinctions ($A_{K_s}$) are ±0.13 and ±0.08 mag. These errors are dominated by scatter of the period-luminosity relations (±0.09 mag) and uncertainties in the reddening law. We reject the possibility that these stars are type II Cepheids. If they were, their distances would be as small as ~2 kpc and in conflict with the large extinctions derived. On the other hand, we also found 17 type II Cepheids located in the Galactic Bulge. The estimated distances and foreground extinctions for these objects reject the possibility that they are classical

Cepheids. The catalogue and characteristics of the whole sample will be given in a future publication (N.M. *et al.*, in preparation).

The three classical Cepheids were also detected by the *Spitzer Space Telescope* although they were not then known to be Cepheids; their IDs in the *Spitzer* photometric catalogue[30] are SSTGC 0596712, 0503948 and 0500191 for (a), (b) and (c) in Table 1, respectively. Although crowding affected the *Spitzer* photometry for object (b), the distances based on the mid-IR period-luminosity relation[31] for the others agree well with those obtained by our near-IR photometry.

The properties of the Cepheids such as periods and locations on the colour-magnitude diagram (Supplementary Fig. 1) are close to each other. In addition, as mentioned in the text, the three Cepheids are located close to the plane containing the central black hole, at galactic latitude of −0.05°, which agrees with the thin disc-like structure of the nuclear bulge[1,5]. The above evidence clearly suggests that all the three objects are classical Cepheids located within the nuclear bulge.

**2. Possible cluster origins of the nuclear bulge Cepheids:**

Two objects, (b) and (c) in Table 1, lie close to each other at least in projection. Whilst it might be possible that they both belonged to a stellar cluster or association, we found no clear concentration of stars surrounding them. Theoretical investigations suggest that stellar clusters like Arches[32] at around the Galactic Centre can be destroyed, or disperse so as to fade into the dense background field, on a short time scale (~10 Myr)[33].

**3. Detection limit of classical Cepheids:**

Shorter-period Cepheids are fainter, so that the completeness limit may affect the observed period distribution. The period-luminosity relation, giving an absolute magnitude, enables us to tell if the detection limit is deep enough to detect a Cepheid at the distance of the nuclear bulge with a given period and foreground extinction. The typical limiting magnitudes of our survey are 16.4, 14.5 and 13.1 mag, while the saturation limit is at 9.5, 9.5 and 9.0 mag in the $J$, $H$ and $K_s$ bands, respectively[9] (see Supplementary Fig. 1). Note that these limits depend on the seeing of each image and the crowdedness across the field[9], thus the typical values are indicated. Supplementary Fig. 2 illustrates the detectable range in the parameter plane of (log $P$, $A_{K_s}$). In the study of Mira-type variable stars towards the Galactic Centre[9], it was shown that a large fraction of the stellar populations had foreground extinctions of 2—3 mag in $K_s$. According to Supplementary Fig. 2, we can detect Cepheids in the nuclear bulge with log $P$ larger than 0.7, at least in the $K_s$ band for the lines of sight with $A_{K_s}$ between 2 and 3 mag. We searched for variable stars by using the three band datasets independently, so that we should have found those visible at least in one band if they existed. Placing hypothetical shorter-period, i.e. fainter, Cepheids than the ones detected on the colour-magnitude diagram (Supplementary Fig. 1) also supports the conclusion that they would have been detected if they were present in the nuclear bulge. We conclude that the absence of Cepheids with 5 days < $P$ < 19 days is not a result of the survey limit.

**4. Estimation of the star formation rate:**

According to evolutionary models of Cepheids[6,21], stars evolving into Cepheids with $P$~20 days have initial masses around 8—10 solar masses (corresponding to ages between 20 and 30 Myr), and their lifetime as Cepheids is about 0.1 Myr[21]. The initial total mass of the parent population needed to explain the presence of three Cepheids can be estimated assuming a suitable initial mass function (hereinafter IMF). We here adopt the IMF suggested in ref. 20, in order to make a direct comparison with the result in ref. 5, and consider the effect of the assumed IMF. The IMF in ref. 20 is close to the "standard" one[34]. Some early papers suggested that the IMF in the Galactic Centre region may be different from "standard" (see ref. 35 for a review), but recent investigations[32,36] suggest that the difference is small if any. Then, a population of $10^5$ solar masses in total is expected for the age range of 20—30 Myr to produce three Cepheids of the relevant periods. This indicates a star formation rate of 0.01 solar masses per year for the corresponding look-back time. The Poisson noise of the Cepheid number (a factor of $3^{1/2}$) and the uncertainty of the Cepheid lifetime (a factor of 2) lead to an error of the estimated star formation rates being a factor of 2.5.

On the other hand, the absence of short-period Cepheids with ages of 30—70 Myr (initial masses between 6 and 8 solar masses) indicates that the stellar population within the age range is smaller than $10^5$ solar masses (estimated as the 1σ upper limit assuming Poisson statistics). This gives a star formation rate less than 0.0025 solar masses per year considering the corresponding Cepheid lifetimes, 0.2 Myr. Note that such an upper limit can be placed independently of the estimate for the younger Cepheids above. With the IMF we adopted[20], we would expect 4.3 Cepheids with 6—8 solar mass if the star formation rate had remained 0.004 solar masses per year (the 1σ lower limit of the star formation rate between 20—30 Myr ago). However, this would lead to a low probability of 1.4 per cent for finding no Cepheids. The change of the star formation rate is significant even though we detected only three Cepheids. A shallower IMF would predict a smaller number of the Cepheids with 6—8 solar mass relative to the number of the more massive Cepheids. With the IMF in ref. 37, the expected number of Cepheids with 6—8 solar mass would become 3.0 in the above example with the constant star formation rate, but the probability for finding no Cepheids is still low ($P < 5$ per cent).

Our survey field does not include the entire nuclear bulge and is smaller than the region considered in ref. 5. Considering the mass distribution within the nuclear bulge[1], we expect about 20 per cent of the whole nuclear bulge stars in our field. In addition, our survey was effective for only a part of the populations (~2/3) considering that the others are obscured behind the extreme foreground extinction. Therefore we apply a correction by a factor of 7.5 for the incompleteness of our survey, and thus obtain 0.075 and 0.02 (1σ upper limit) solar masses per year for 20—30 Myr and 30—70 Myr ago respectively. In Fig. 3, the relevant previous estimates[4,5] are also included. The star formation rates in ref. 5 were obtained for young stellar objects with 4.5 μm excess, Stage I young stellar objects, 24 μm flux density representative of the ionizing flux and non-thermal radio emission in the order of look-back time. In contrast, it is suggested in ref. 4 that a continuous star formation rate fits the observed luminosity function better than scenarios with an intensive burst. Their rate obtained for a fixed stellar mass of $2\times10^8$ solar masses is converted to the rate for the entire nuclear bulge with the total mass of $(1.4\pm0.6)\times10^9$ solar masses[1], ~0.14 solar masses per year. Our conversion here is affected by the size of the uncertainty in the total mass, ~60 per cent, which roughly corresponds to the width of the blue stripe in Fig. 3.

Some bright red supergiants were considered to be in the age range discussed here[3], but the age estimates for such objects are less certain than those for Cepheids[38]. Some of the brightest late-type stars may be as young as the three Cepheids, and others may be older than 70 Myr although we found no evidence for or against such an older population since the corresponding short-period Cepheids of such an age would be below our detection limit.

**Supplementary references:**

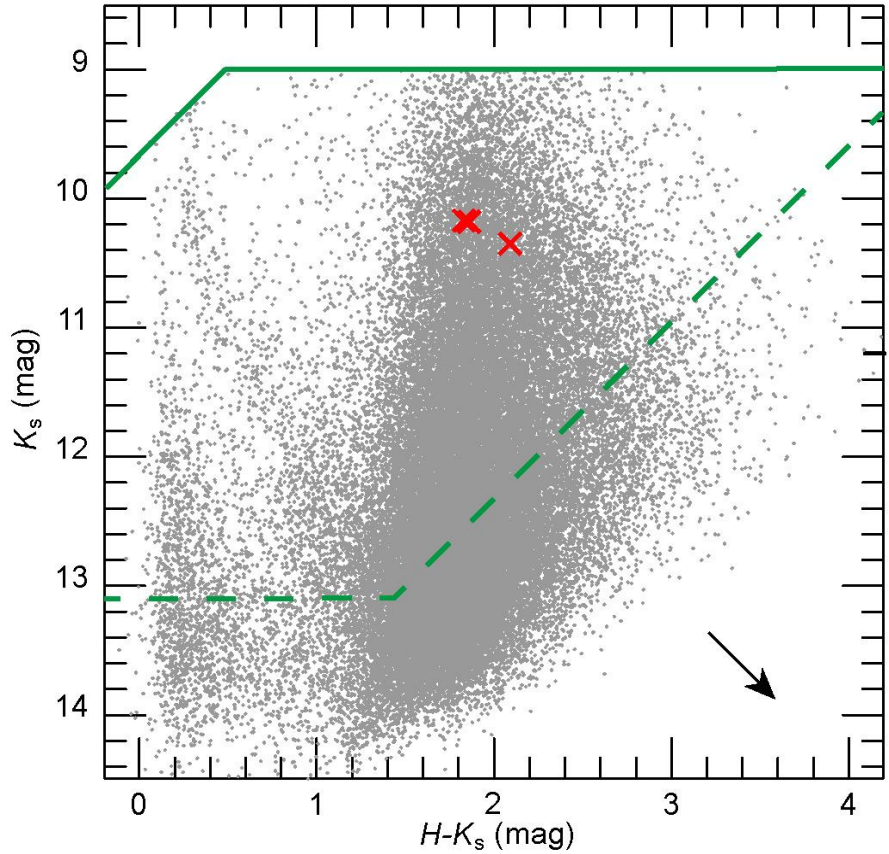

**Supplementary Figure 1 | The colour-magnitude diagram of the classical Cepheids and other stars within our survey field.**

On the colour-magnitude diagram, $H$–$K_s$ versus $K_s$, the classical Cepheids are indicated by red crosses, while other stars are indicated by grey dots. The effect of interstellar extinction is indicated by the arrow whose length corresponds to an extinction of 0.5 mag in $K_s$. The filled green curve indicates the saturation limit, and the dashed curve indicates the detection limit. The following three groups of stars are prominent. The almost vertical sequence on the left hand side is the main sequence of the foreground. Diagonally running from ($H$–$K_s$=0.3, $K_s$=10) down to ($H$–$K_s$=1, $K_s$=12.5) is the sequence of foreground red-clump giants whose reddenings get larger with the increasing distance. Most sources belong to the reddened group, $H$–$K_s$>1.5, which includes red giants and other types of bright stars near the Galactic Centre. The three classical Cepheids belong to this group, and their colours and magnitudes are similar to each other.

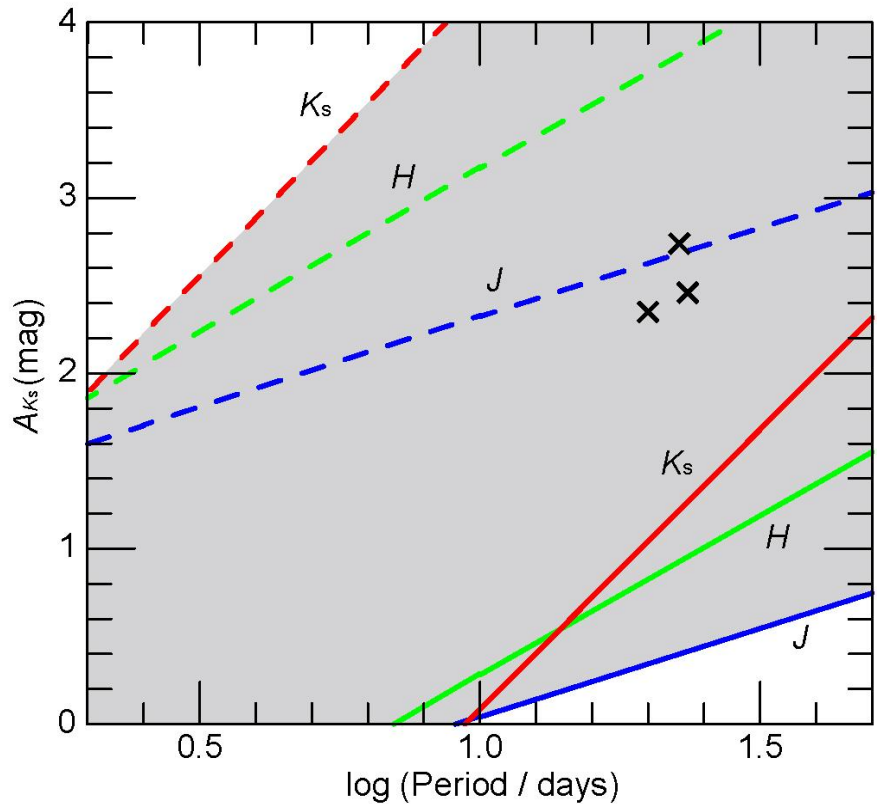

**Supplementary Figure 2 | A schematic diagram showing the range where classical Cepheids could be detected.**

This diagram illustrates the parameter space of the pulsation period (log $P$) and the $K_s$-band foreground extinction ($A_{K_s}$) in which we can detect classical Cepheids located at the distance of the nuclear bulge. Dashed lines show the detection limits for $JHK_s$, while solid lines show the saturation limits. The cross symbols indicate the observed parameters of the detected Cepheids. The Cepheids with the typical extinctions at the nuclear bulge, $2 < A_{K_s} < 3$ (ref. 9), would be above the detection limits, at least in the $K_s$ band, if their periods were longer than 5 days as indicated with the grey region.